\documentclass[aps,prl,twocolumn,groupedaddress]{revtex4-1}
\usepackage{graphicx}
\usepackage{amsmath}
\usepackage{dcolumn}
\usepackage{bm}

\begin{document}

\preprint{}

\title{Magneto-reflection spectroscopy of monolayer transition-metal dichalcogenide semiconductors in pulsed magnetic fields}

\author{Andreas V. Stier$^1$, Kathleen M. McCreary$^2$, Berend T. Jonker$^2$, Junichiro Kono$^3$, Scott A. Crooker$^1$}

\affiliation{$^1$National High Magnetic Field Laboratory, Los Alamos, New Mexico 87545, USA}
\affiliation{$^2$Materials Science and Technology Division, Naval Research Laboratory, Washington, DC 20375, USA}
\affiliation{$^3$Department of Electrical and Computer Engineering, Department of Physics and Astronomy, and Department of Materials Science and NanoEngineering, Rice University, Houston, Texas 77005, USA}

\begin{abstract}

We describe recent experimental efforts to perform polarization-resolved optical spectroscopy of monolayer transition-metal dichalcogenide semiconductors in very large pulsed magnetic fields to 65 tesla. The experimental setup and technical challenges are discussed in detail, and temperature-dependent magneto-reflection spectra from atomically thin tungsten disulphide (WS$_2$) are presented. The data clearly reveal not only the valley Zeeman effect in these 2D semiconductors, but also the small quadratic exciton diamagnetic shift from which the very small exciton size can be directly inferred. Finally, we present model calculations that demonstrate how the measured diamagnetic shifts can be used to constrain estimates of the exciton binding energy in this new family of monolayer semiconductors.
\end{abstract}

\maketitle

\section{I. INTRODUCTION}

Historically, magneto-optical studies have played a central role in revealing the fundamental properties of excitons in bulk and low-dimensional semiconductors. Various polarization-resolved optical spectroscopies in applied magnetic fields have helped to determine the mass, size, energy, magnetic moment, and dimensionality of excitons and carriers in a great many conventional semiconductor materials \cite{Miura, Knox}.  Recently, a new family of atomically-thin semiconductors known as the monolayer transition-metal dichalcogenides (TMDs) has captured the attention of physicists, materials scientists, and chemists working broadly in the fields of semiconductors and 2D materials \cite{Splendiani, Mak2010}. These new monolayer TMDs, which include atomically-thin flakes and films of MoS$_2$, MoSe$_2$, WS$_2$ and WSe$_2$, are semiconductors possessing direct optical bandgaps at the $K$ and $K'$ points of their hexagonal Brillouin zone. Owing to strong spin-orbit coupling and their lack of structural inversion symmetry, spin and valley degrees of freedom are coupled and \emph{valley-specific} optical selection rules exist for right- and left- circularly polarized light \cite{Xiao, XuReview}. Consequently, a number of interesting optical and magneto-optical studies of these TMDs have been performed in recent years, in which both spin and valley physics were explored \cite{MakNatNano, Zeng, Sallen, Cao, MacNeill, Srivastava, Aivazian, Li, Wang, Mitioglu, Yang1, Yang2, Stier}.

On both theoretical and experimental grounds \cite{Ram, Komsa, Liu, Ye, KormanyosPRX, Chernikov, He, Zhu, WangPRL2015, Hanbicki, Stroucken}, electron and hole masses in the monolayer TMDs are thought to be rather heavy (of order $0.5 m_0$, where $m_0$ is the free electron mass), the exciton binding energies are reported to be extremely large (of order 500-1000~meV), and the physical sizes of the excitons are predicted to be very small (of order 1-2~nm).  [In comparison, in GaAs the electron mass is 0.067$m_0$, the exciton binding energy is only 4~meV, and the exciton Bohr radius is $\sim$20~nm.] For these reasons -- and also because the photoluminescence and absorption linewidths in monolayer TMDs are relatively broad ($\sim 10-40$~meV, depending on the material) -- very large magnetic fields of order 50-100~T are desirable so that the small Zeeman shifts and the even-smaller exciton diamagnetic shifts can be clearly resolved in experimental data.

To this end we have recently developed capabilities for performing polarization-resolved magneto-reflection studies of monolayer TMD materials at cryogenic temperatures down to 4~K and in very high pulsed magnetic fields to 65~T. We recently reported the first results of such measurements (on monolayer MoS$_2$ and WS$_2$) in Ref. \cite{Stier}. The intent of this paper is therefore to present a considerably more detailed description of the experimental setup and of the challenges faced when working with monolayer materials in pulsed magnetic fields. In particular we focus on how we achieve and verify the circular polarization selectivity, and how we mitigate problems due to the mechanical vibrations that are ubiquitous in pulsed-field studies. We present new data showing temperature-dependent studies of the valley Zeeman effect and exciton diamagnetic shift.  Finally, we also extend recent calculations of the exciton binding energy in these monolayer TMDs to a more realistic case that includes the effect of the dielectric substrate, and discuss the experimental results within that context.

\section{II. EXPERIMENT}

\subsection{A. Setup}

Figure 1(a) shows one of the 65~T pulsed magnets used at the National High Magnetic Field Laboratory (NHMFL) at Los Alamos National Laboratory. The magnet has a 15~mm bore and is powered by a 16~kV, 32~mF capacitor bank.  A liquid helium bath cryostat sits atop the magnet, and has a long vacuum-insulated tail section that extends into the magnet bore. During operation, the magnet is immersed in liquid nitrogen to reduce its initial resistance. Full-field pulses can be repeated every 45 minutes, limited by the cool-down time of the magnet following each pulse. A representative field profile from this magnet is shown in Fig. 1(b).  Magneto-reflectance studies were performed with the samples at cryogenic temperatures down to 4~K using a home-built fiber-coupled optical probe depicted in Fig. 1(c). The probe, which resides in an additional vacuum jacket filled with helium exchange gas, is constructed from nonmetallic fiberglass (G10) and polycarbonate materials (Vespel) and has a diameter of 8~mm.

Broadband white light from a xenon lamp was coupled to the samples using a 100~$\mu$m diameter multimode optical fiber. The light was focused onto the sample at near-normal incidence using a single aspheric lens (6~mm focal length, NA 0.3), and the reflected light was refocused by the same lens into a 600~$\mu$m diameter collection fiber.  The collected light was dispersed in a 300~mm spectrometer and was detected with a liquid nitrogen cooled charge-coupled device (CCD) detector. The CCD was configured to acquire full spectra continuously at a rate of about 500~Hz ($\sim$2~ms/spectra) throughout the magnet pulse. The blue spikes in Fig. 1(b) show the CCD timing signal. The choice of 100~$\mu$m diameter for the delivery fiber achieves a satisfactory balance between the conflicting goals of achieving a small focused spot on the sample whilst still allowing a sufficient amount of light to be coupled from the xenon lamp to the sample.

Polarization selectivity is achieved via a thin-film circular polarizer that can be mounted over either the delivery fiber or the collection fiber. Depending on the configuration and on the direction of the magnetic field (positive or negative), this provides sensitivity to the $\sigma^+$ or $\sigma^-$ polarized optical transitions in the $K$ or $K'$ valleys of monolayer TMDs, as discussed in more detail below.

Figure 1(d) shows a typical image of the large-area monolayer WS$_2$ films used in these experiments.  These films are grown by chemical vapor deposition (CVD) on Si/SiO$_2$ substrates and typically have mm-square regions with $>$99\% monolayer coverage \cite{McCreary}, which is much larger than the $\sim$100~$\mu$m diameter spot of the focused white light on the sample.  Details of the sample growth and characterization of film quality from photoluminescence and Raman spectroscopy can be found in Refs. \cite{McCreary, Stier}.

The use of large-area CVD-grown films was critical for these pulsed magnetic field experiments. While stable optical alignment onto micron-scale exfoliated TMD flakes is relatively straightforward in low-field superconducting and dc magnets \cite{Srivastava, MacNeill, Aivazian, Li, Wang}, the mechanical vibrations inherent in pulsed magnets significantly complicate such approaches. Moreover, the in-situ nanopositioners commonly used in dc magneto-spectroscopy of micron-scale samples are generally not amenable to the small bore sizes and rapidly varying magnetic field environment of a high-field pulsed magnet. These stringent alignment requirements are considerably relaxed, however, when using larger-area samples having high spatial uniformity because micron-scale vibrations and temperature drifts do not affect the detected signals to leading order. In this case, a fiber-coupled probe of the type described above typically suffices to obtain high quality spectra that are largely free from mechanical vibrations and subsequent misalignment during the magnet pulse. Similar fiber-coupled probe designs have been successfully used in conjunction with pulsed magnets to study mm-squared samples of magnetic semiconductors \cite{Crooker1, Crooker2}, quantum wells \cite{Astakhov, Alberi}, colloidal quantum dots \cite{Zeke, Furis}, carbon nanotubes \cite{Zaric, Shaver}, and polymers \cite{Diaconu}.

\begin{figure}[tbp]
\center
\includegraphics[width=.40\textwidth]{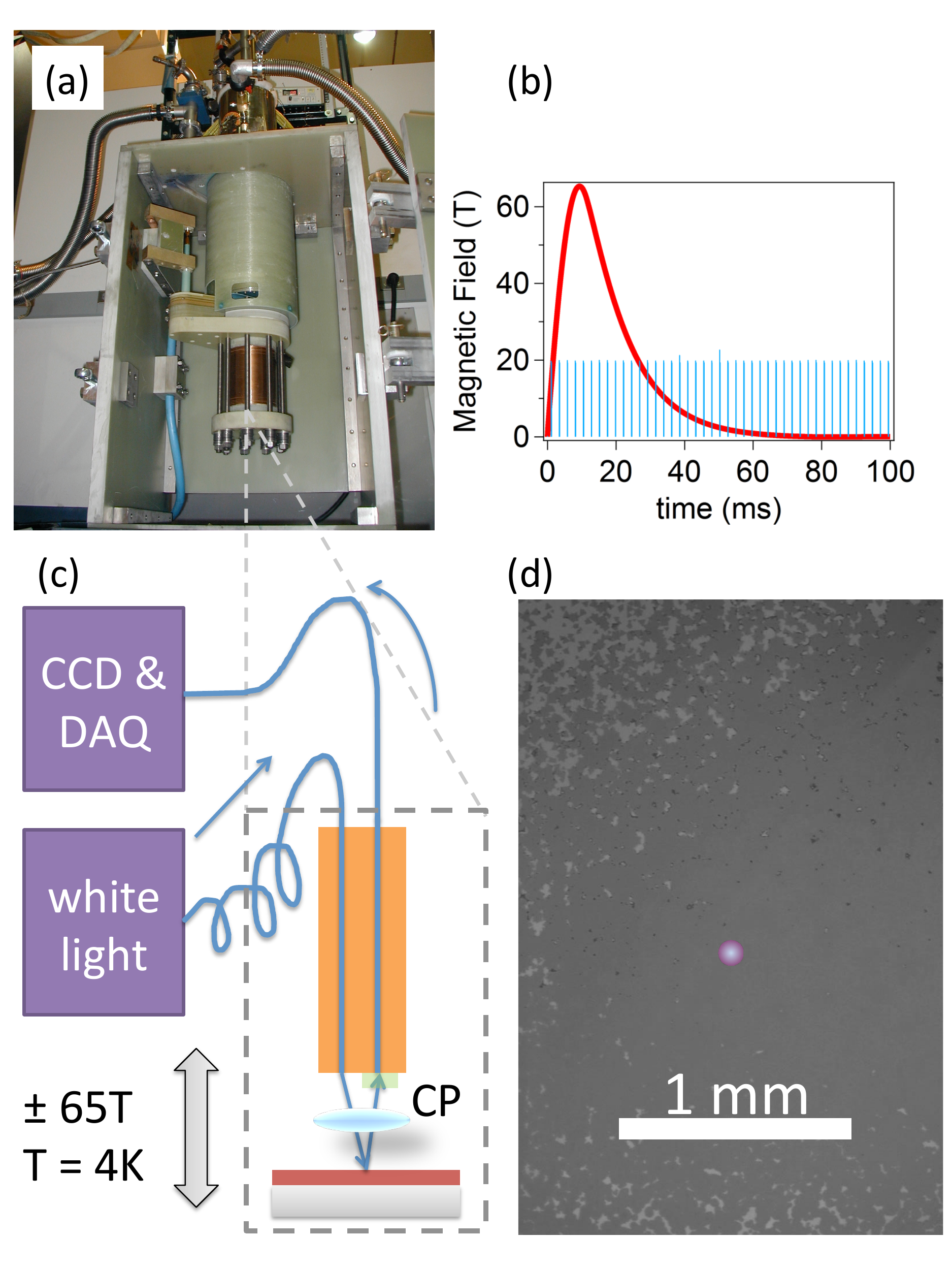}
\caption{(a) Photograph of a 65~T capacitor-driven pulsed magnet at the NHMFL. The tail of a liquid helium bath cryostat extends down into the magnet bore.  (b) Typical field profile (red curve). The blue spikes show the timing signal of the CCD; full 16-bit optical spectra are acquired at each spike. (c) Schematic of the fiber-coupled optical reflection probe. White light is coupled to the sample via a 100~$\mu$m diameter optical fiber and a single aspheric lens. Reflected light is refocused back into an adjacent 600~$\mu$m diameter collection fiber, and detected with a spectrometer and CCD. (d) Optical microscope image of a large-area WS$_2$ film grown by CVD on a Si/SiO$_2$ substrate. Dark (bright) regions show the monolayer WS$_2$ film (substrate). The spot size of the focused white light is $\sim$100~$\mu$m as indicated by the red circle.} \label{fig1}
\end{figure}

\subsection{B. Excitons in monolayer WS$_2$}
Figure 2(a) shows the low-temperature reflection spectrum from a monolayer WS$_2$ film at zero magnetic field. The A and B exciton transitions are clearly visible. Their origin can be understood from Figure 2(b), which depicts the conduction and valence bands at the $K$ and $K'$ valleys, as well as the associated exciton transitions and optical selection rules. Strong spin-orbit coupling of the valence band splits the spin-up and spin-down components by $\sim$400~meV in monolayer WS$_2$, giving rise to the large separation between the A and B exciton transitions. Owing to the valley-specific optical selection rules in these monolayer TMD materials, $\sigma^+$ circularly-polarized light couples to both A and B exciton transitions in the $K$ valley, while light of the opposite $\sigma^-$ circular polarization couples to the exciton transitions in the $K'$ valley.

At zero applied magnetic field, the bands and optical transitions in the $K$ and $K'$ valleys are nominally degenerate in energy and related by time-reversal symmetry. That is, spin-up conduction (valence) bands in $K$ and spin-down conduction (valence) bands in $K'$ have the same energy and equal-but-opposite total magnetic moment ($\bm{\mu}_K^{\textrm{c,v}} = - \bm{\mu}_{K'}^{\textrm{c,v}}$). Therefore an applied magnetic field, which breaks time-reversal symmetry, will lift the $K/K'$ valley degeneracy by shifting time-reversed pairs of states in opposite directions in accord with the Zeeman energy $-\bm{\mu} \cdot \textbf{B}$. This will Zeeman-shift the measured exciton energy if the relevant conduction and valence band moments are \emph{unequal}; $\Delta E_\textrm{Z} = -(\bm{\mu}^\textrm{c} - \bm\mu^\textrm{v})\cdot \textbf{B}$.

\begin{figure}[tbp]
\center
\includegraphics[width=.40\textwidth]{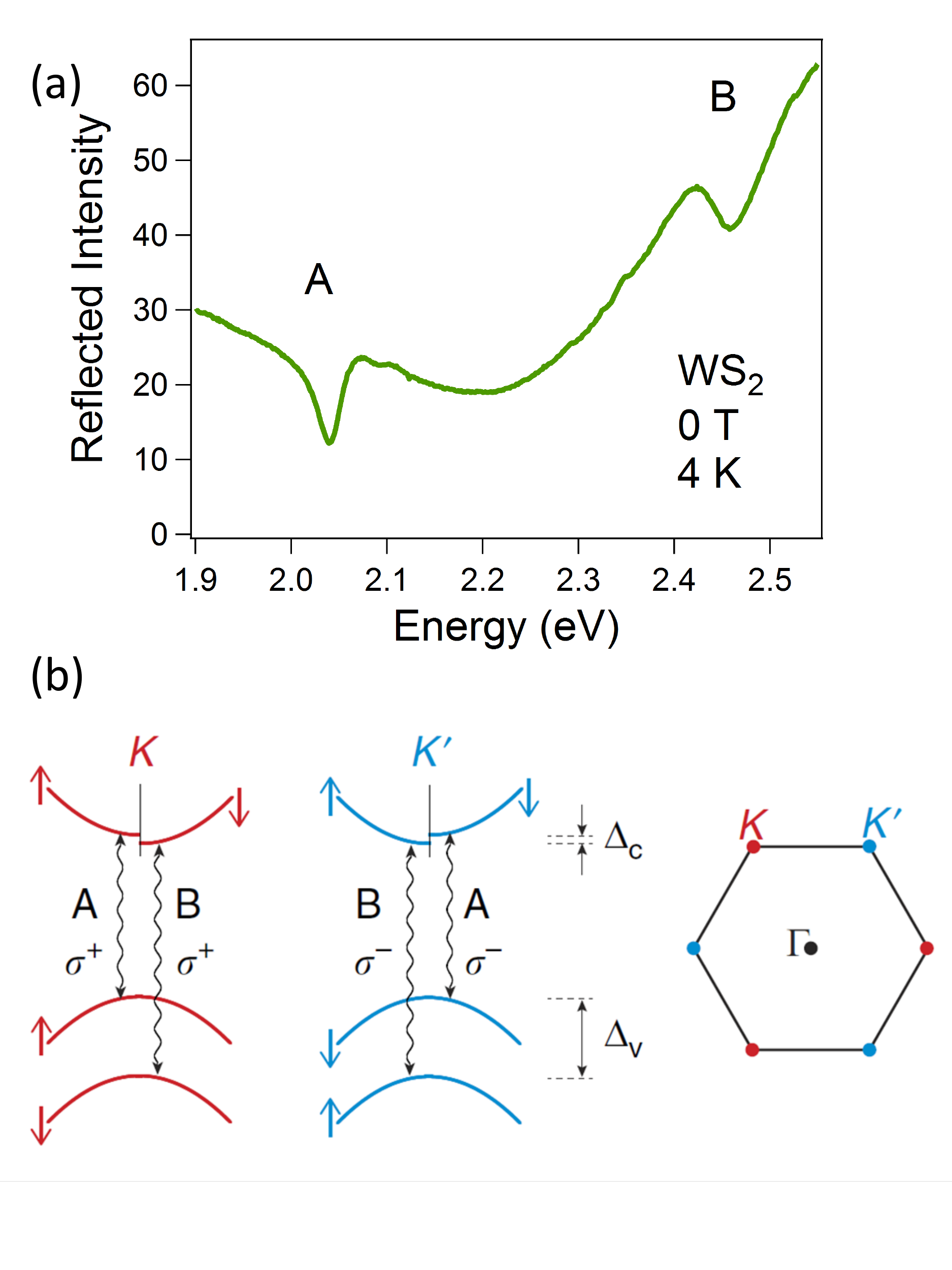}
\caption{(a) Reflection spectrum from a monolayer WS$_2$ film at zero magnetic field and at $T = 4$~K. The A and B exciton features are clearly visible on top of a smoothly varying background. (b) Schematic of the conduction and valence bands in the vicinity of the \emph{K} and \emph{K'} valleys of monolayer WS$_2$. The A and B exciton optical transitions (wavy lines) and the associated optical selection rules for circularly-polarized $\sigma^+$ and $\sigma^-$ light are indicated. Spin-orbit coupling splits the spin-up and spin-down states in the conduction and valence bands ($\Delta_\textrm{c} \approx $30~meV, $\Delta_\textrm{v} \approx $400~meV).} \label{fig2}
\end{figure}

\subsection{C. Valley Zeeman Effect}

\begin{figure}[tbp]
\center
\includegraphics[width=.48\textwidth]{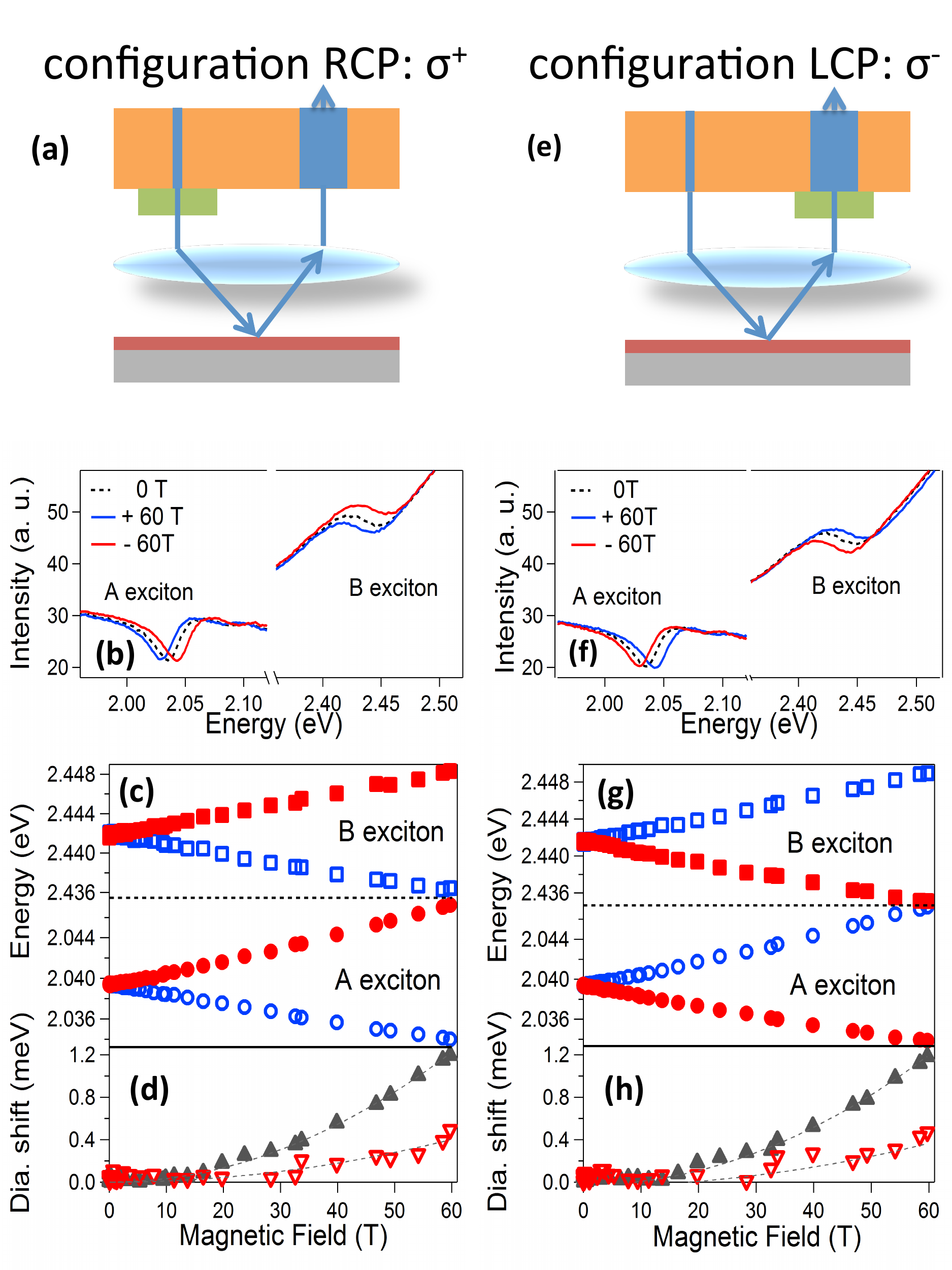}
\caption{Comparison of the magneto-reflection signals for the two experimental configurations: The circular polarizer film (green) positioned over the delivery fiber, or over the collection fiber, respectively. Blue (red) data are for positive (negative) magnetic fields. Black dashed data are for zero field. $T$=4~K. (a,b) Magneto-reflection spectra of the A and B excitons for the first configuration. The blue trace was acquired at $+60$~T and corresponds to the $\sigma^+$ transitions in the \emph{K} valley. The red trace at $-60$~T is equivalent (by time-reversal symmetry) to the $\sigma^-$ transitions in the \emph{K'} valley. The valley Zeeman splitting of the optical transitions is clearly visible. (c) Energies of the field-split A and B excitons versus magnetic field. (d) The \emph{average} energy of the field-split exciton peaks reveals the small quadratic diamagnetic shift of the exciton. (e)-(f) Similar data and analysis for the second experimental configuration, wherein the circular polarizing film is positioned over the collection fiber.} \label{fig3}
\end{figure}

As described above, we selectively probe transitions in the $K$ or $K'$ valley by using a circular polarizer film (linear polarizer + quarter wave plate) mounted either over the delivery fiber or over the collection fiber.  Since it is difficult to switch the position of the polarizer during an experiment (this would require disassembly of the probe and would likely lead to a different spot that is probed on the sample), we typically fix the position of the circular polarizer (\emph{e.g.} over the delivery fiber), and pulse the magnet in the positive ($+65$~T) and then the negative ($-65$~T) field direction. For nonmagnetic samples, the latter case is in principle equivalent (by time-reversal symmetry) to measuring the $\sigma^-$ optical transitions in \emph{positive} field.  Nonetheless, we did verify that measurements using both configurations of the circular polarizer gave consistent results, as shown in Fig. 3.  Absolute sign conventions were confirmed via magneto-reflectance studies of a diluted magnetic semiconductor (Zn$_{.92}$Mn$_{.08}$Se), for which $\sigma^\pm$ optical transitions are easily identified in small magnetic fields.

Figure 3 shows circularly polarized magneto-reflection spectra from monolayer WS$_2$ in pulsed fields to $\pm 60$~T using both configurations of the circular polarizing film.  The expected symmetry between positive and negative magnetic fields is confirmed (blue and red data traces, respectively). The spectra reveal a well-resolved splitting of the A and B exciton of $\sim$14 meV at 60~T, and the derived \emph{g}-factors of approximately $-4$ agree well with our recently published results \cite{Stier} and are in reasonable agreement with recent reports of the valley Zeeman effect in the monolayer transition-metal diselenides WSe$_2$ and MoSe$_2$ \cite{Li, MacNeill, Aivazian, Srivastava, Wang, Mitioglu}.

These measurements (here and in Ref. \cite{Stier}) provide the first experimental values of the valley Zeeman effect of both the A \emph{and} B excitons in monolayer TMD materials. As discussed in detail in Ref. \cite{Stier}, the fact that the measured values of $g \simeq -4$ for \emph{both} A and B excitons is unexpected and surprising, because the reduced mass of these two excitons are expected to be different (and indeed, our data provide strong experimental evidence for a mass difference, as described below).

\subsection{D. Exciton Diamagnetic Shift}

The use of very large magnetic fields also permits the first observation of the small quadratic \emph{diamagnetic shift} of excitons in these monolayer TMD materials.  The exciton diamagnetic shift is a fundamental and very important parameter in semiconductor physics \cite{Miura, Knox, Walck}, because it allows to directly measure the physical size of the exciton -- an essential material parameter.

Independent of the functional form of the Coulomb potential and the resulting shape of the exciton wavefunction \cite{Miura, Knox, Walck}, an exciton's diamagnetic shift $\Delta E_{\rm dia}$ is given by
\begin{equation}
\Delta E_{\rm dia} = \frac{e^2}{8 m_r} \langle r^2 \rangle B^2 = \sigma B^2,
\end{equation}
where $\sigma$ is the diamagnetic shift coefficient, $m_r$ is the exciton reduced mass, $r$ is a radial coordinate in a plane perpendicular to the applied magnetic field $B$, and $\langle r^2 \rangle = \langle \psi_{\rm 1s}| (x^2 + y^2) |\psi_{\rm 1s} \rangle$ is the expectation value of $r^2$ over the 1s exciton wavefunction $\psi_{\rm 1s}(r)$. Equation (1) applies in the `low-field' limit where $\Delta E_{\rm dia}$ and the cyclotron energy $\hbar \omega_c$ are less than the exciton binding energy, which is still very much the case in the monolayer TMDs even at 65~T. Given $m_r$, $\sigma$ can then be used to determine the root-mean-square (rms) radius of the 1s exciton in the monolayer plane, $r_1$:
\begin{equation}
r_1 = \sqrt{\langle r^2 \rangle_{\rm 1s}} = \sqrt{8 m_r \sigma}/e.
\end{equation}

Note that $r_1$ is not the conventionally-defined exciton Bohr radius $a_0$. The notion of a Bohr radius applies to classic Coulomb potentials that scale as $-1/r$, for which $a_0$ appears in the functional form of the exciton wavefunction $\psi (r)$. As described below, such conventional potentials likely do not apply to real 2D materials. The rms exciton radius $r_1$ is a well-defined parameter for any arbitrary exciton wavefunction (in a conventional bulk material where $V(r) \propto -1/r$, $r_1 = \sqrt{2} a_0$).

To directly reveal $\Delta E_{\rm dia}$, Figs. 3(d) and 3(h) show the \emph{average} exciton energy versus magnetic field. Overall quadratic shifts are observed, indicating diamagnetic coefficients $\sigma_\textrm{A}= 0.32~\mu$eV/T$^2$ for the A exciton and  $\sigma_\textrm{B} = 0.11~\mu$eV/T$^2$ for the B exciton (independent of the circular polarizer configuration, as expected). To infer the exciton radius $r_1$, an exciton reduced mass $m_r$ must be assumed. Theoretical estimates \cite{Xiao, Berkelbach, Ram} for the A exciton reduced mass in monolayer WS$_2$ range from $m_{r,\textrm{A}} = 0.15-0.22 m_0$, which allow us to directly calculate $r_{1, \textrm{A}}=1.48-1.79$~nm via Eq.~(2). These values of $r_1$ are in reasonable agreement with recent \emph{ab initio} calculations of the 1s exciton wavefunction in monolayer WS$_2$ \cite{Ye}, and further support a picture of Wannier-type excitons with lateral extent larger than the monolayer thickness (0.6~nm) and spanning several in-plane lattice constants. These results were discussed in detail in Ref. \cite{Stier}.

\begin{figure}[tbp]
\center
\includegraphics[width=.48\textwidth]{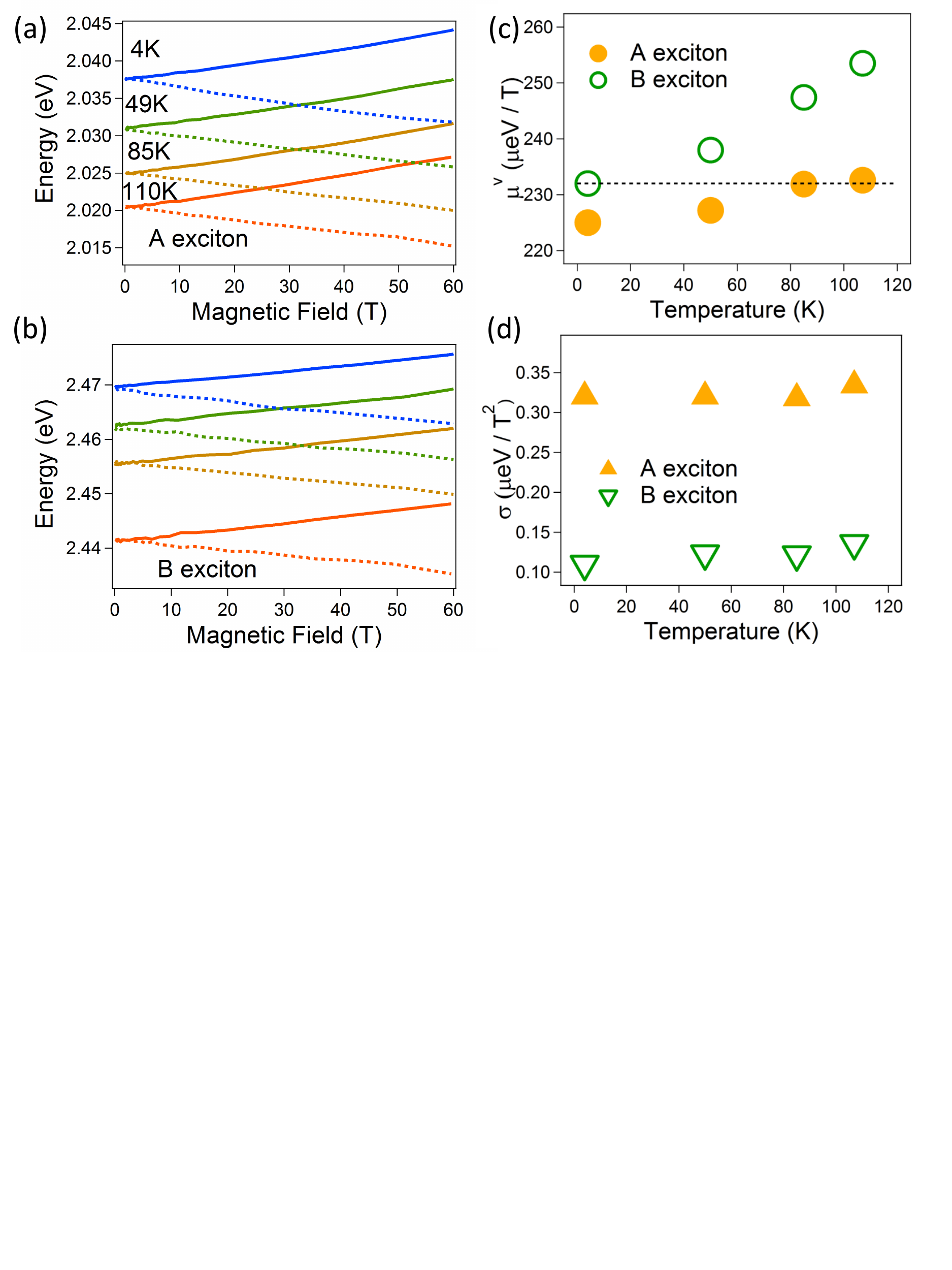}
\caption{(a) Energy of the A exciton versus magnetic field in monolayer WS$_2$ for different temperatures. Solid (dashed) lines correspond to positive (negative) magnetic fields. (b) Same, for the B exciton. (c) Valley Zeeman splitting of the A and B excitons versus temperature. (d) Diamagnetic shift of both excitons versus temperature.} \label{fig4}
\end{figure}

Temperature-dependent magneto-reflectivity studies were also performed and are shown in Fig. 4.  Figures 4(a,b) show the A and B exciton energies versus magnetic field (for both $\sigma^+$ and $\sigma^-$ polarizations) at different temperatures up to 110~K. The zero-field exciton energies redshift with increasing temperature, as previously reported.  The valley Zeeman splitting for the A exciton is largely unchanged with temperature while that for the B exciton slightly increases [Fig. 4(c)]. Importantly, the diamagnetic shift, and therefore the size of the excitons, remains essentially unchanged with increasing temperature [Fig. 4(d)], indicating that the observed temperature-dependent red-shift of the exciton energy is likely due to the reduction of the single-particle bandgap of the material and not to any significant change in exciton properties.

\section{III. DISCUSSION}

Finally, we discuss how knowledge of the exciton diamagnetic shift can also be used to place constraints on estimates of the exciton binding energy -- a parameter of significant current interest in the monolayer TMDs, for which both experimental and theoretical estimates vary considerably \cite{Ram, Komsa, Liu, Ye, KormanyosPRX, Chernikov, He, Zhu, WangPRL2015, Hanbicki, Stroucken}.  This procedure was discussed in Ref. \cite{Stier} for the case of suspended monolayer TMDs, and here we briefly review these arguments and also present new simulations for the more realistic case of monolayer TMDs on a dielectric substrate.

In contrast to bulk materials, the attractive electrostatic potential $V(r)$ between an electron and hole in 2D materials does not scale simply as $1/r$.  This is due the phenomenon of \emph{non-local dielectric screening}, wherein the effective dielectric constant that is `seen' by an exciton depends strongly on the electron-hole separation \cite{Keldysh, Cudazzo, Berkelbach, Chernikov}. Rather, the potential $V(r)$ in a free-standing 2D material in vacuum is believed to assume the following form:
\begin{equation}
V(r)=-\frac{e^2}{8 \varepsilon_0 r_0}\left[ H_0 \left(\frac{r}{r_0} \right) - Y_0 \left(\frac{r}{r_0}\right) \right],
\end{equation}
where $H_0$ and $Y_0$ are the Struve function and the Bessel function of the second kind, respectively, and the characteristic length scale $r_0$ is the screening length $r_0 = 2 \pi \chi_{\rm 2D}$, where $\chi_{\rm 2D}$ is the 2D polarizability of the monolayer material. This potential approaches the classic $-1/r$ form at large electron-hole separations ($r \gg r_0$), but diverges only weakly as log($r$) for small $r$, leading ultimately to a markedly different Rydberg series of exciton states with considerably modified wavefunctions and binding energies \cite{Berkelbach, Chernikov}.

Using this potential, it is possible to numerically calculate via the Schroedinger equation the lowest 1s exciton wavefunction $\psi_{\textrm{1s}} (r)$ and its binding energy for any given reduced mass $m_r$ and screening length $r_0$.  Figure 5(a) shows a color surface plot of the calculated exciton binding energy over a range of possible $m_r$ and $r_0$. Different colors indicate different binding energies.  In addition to the binding energy, the rms exciton radius $r_1$ and also the expected diamagnetic shift $\sigma$ were also calculated via Equation 2 at each $(m_r, r_0)$ point.

Crucially, the solid black lines that are superimposed on this plot indicate \emph{contours of constant diamagnetic shift} that correspond to our experimentally-measured values of $\sigma$ for both the A and B excitons in monolayer WS$_2$.  Therefore, within this model, once a particular reduced mass is assumed, then the value of the exciton binding energy can be uniquely determined by the diamagnetic shift.  For example, if the A exciton reduced mass is $m_r = 0.16$ (a common assumption in monolayer WS$_2$), then the $\sigma_{\textrm{A}}$ contour is intercepted at a binding energy of $\sim$410~meV. Moreover, assuming that the dielectric environment and the screening length $r_0$ are similar for the B exciton, then the parameters for the B exciton are located at the point on the $\sigma_{\textrm{B}}$ contour that lies directly to the right of that for the A exciton. This gives a B exciton reduced mass of 0.27$m_0$, a radius of 1.16~nm, and a binding energy of 470~meV. The diamagnetic shift is therefore a very useful parameter that can be used to benchmark theoretical calculations of the exciton size and its relation to the exciton binding energy in various 2D materials.

In this paper, we extend these calculations to include the effect of a substrate having dielectric constant $\varepsilon_s$, which is of course a very common experimental situation. The modified potential now assumes the form \cite{Keldysh, Cudazzo, Berkelbach, Chernikov}:

\begin{equation}
V(r)=-\frac{e^2}{8 \varepsilon_0 r_0}\left[ H_0 \left(\frac{(1+\varepsilon_s) r}{2 r_0} \right) - Y_0 \left(\frac{(1+\varepsilon_s)r}{2 r_0}\right) \right].
\end{equation}

Figure 5(b) shows the modified surface plot of the exciton binding energy for the case $\varepsilon_s = 4$, which approximately corresponds to the dielectric constant of a SiO$_2$/Si substrate. For any given ($m_r, r_0$) pair, the binding energy is reduced as compared to the case of a suspended TMD film where $\varepsilon_s = 1$. Again assuming that $m_r = 0.16$ for the A exciton, now we find that the binding energy is constrained by the $\sigma_{\textrm{A}}$ contour to be $\sim$290~meV, which is near the value reported recently in Ref. \cite{Chernikov} (320~meV) but is significantly less than most theoretical predictions and the values of $\sim$700~meV reported from two-photon excitation studies of monolayer WS$_2$. Importantly, note that in this case where $\varepsilon_s = 4$ the intersection of the $\sigma_{\textrm{A}}$ contour with a vertical line representing $m_r=0.16$ also corresponds quite well to a screening length of $r_0 =3.8$~nm, which is predicted to be the correct value for monolayer WS$_2$ \cite{Berkelbach}.  A systematic experimental study of the diamagnetic shift as a function of substrate material will shed considerable light on the validity and utility of these numerical models.

\begin{figure}[tbp]
\center
\includegraphics[width=.43\textwidth]{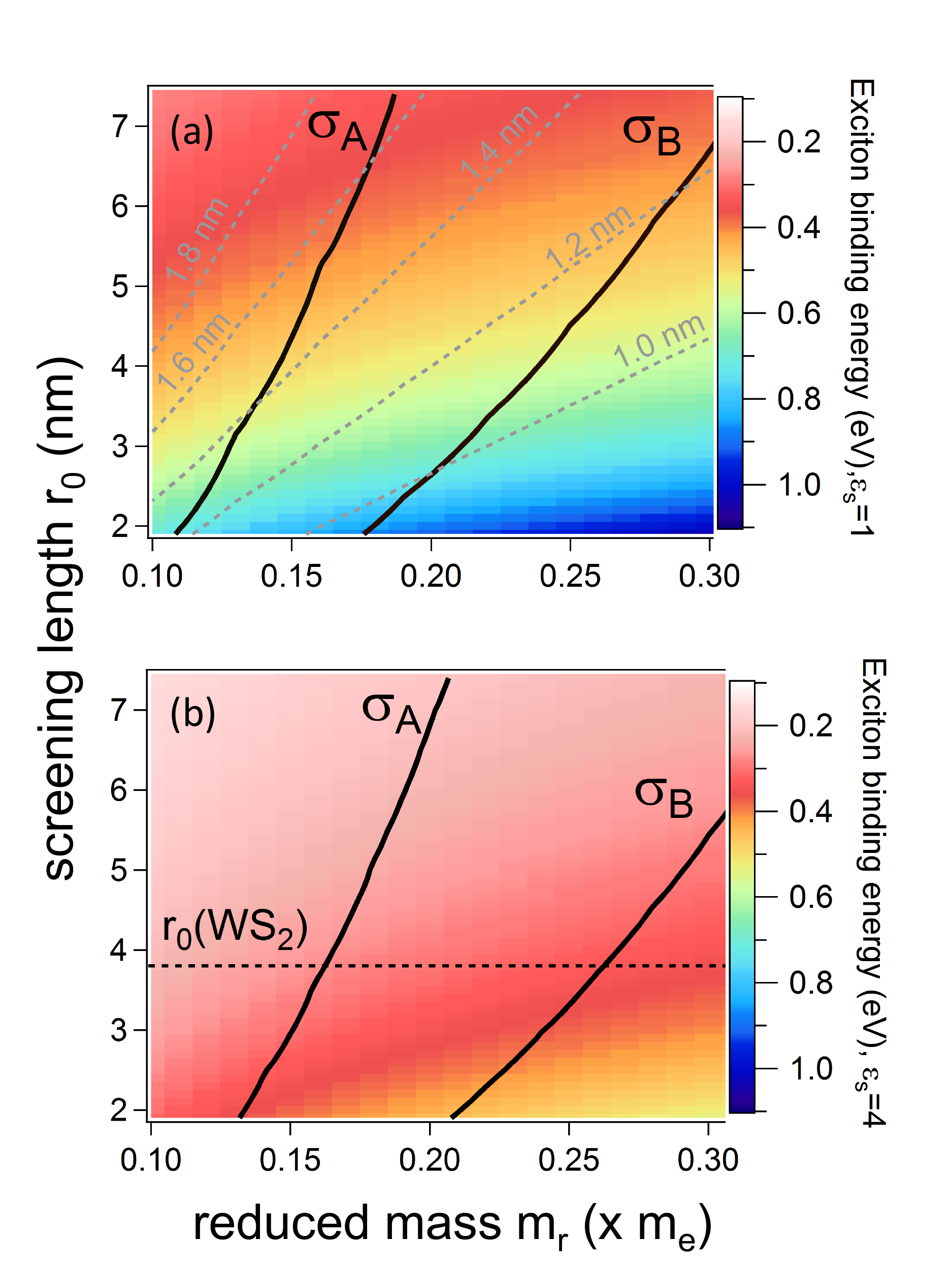}
\caption{(a) Color surface plot of the calculated 1s exciton binding energy using the non-local screening potential $V(r)$ from Equation (3), over a range of possible reduced masses $m_r$ and screening lengths $r_0$. This calculation corresponds to the case of a freestanding (suspended) monolayer TMD film in vacuum. Solid black lines indicate contours of constant diamagnetic shift corresponding to that measured for the A and B excitons in monolayer WS$_2$. Grey dashed lines are contours of constant rms exciton radius $r_1$. (b) Same, but for the case of a monolayer TMD on a substrate with dielectric constant $\varepsilon_s = 4$, which corresponds approximately to Si/SiO$_2$ (see Equation 4). The expected screening length $r_0$ for WS$_2$ (3.8~nm) is indicated by the dashed horizontal line.} \label{fig5}
\end{figure}

In summary, we have described how polarization-resolved optical spectroscopy in very high (pulsed) magnetic fields can reveal new and important parameters of excitons in the recently-discovered family of monolayer TMD semiconductors. Owing to the large masses, small exciton radii, and large exciton binding energies in these 2D materials, pulsed magnetic fields have proven to be an invaluable resource.  Not only can the Zeeman splitting between the broad absorption lines in the monolayer disulphides (\emph{i.e.}, WS$_2$ and MoS$_2$) be clearly resolved -- thereby allowing precise measurements of the valley Zeeman effect -- but the very small quadratic diamagnetic shift of excitons can now be revealed in these 2D materials for the first time. The importance of diamagnetic shift studies is that it allows direct access to the physical size of the excitons, and further, permits one to constrain estimates of the exciton binding energy. Magneto-optical spectroscopy in pulsed magnetic fields is therefore demonstrated to be a powerful tool for the study and characterization of new 2D materials.

\section{ACKNOWLEDGEMENTS}
These measurements were performed at the NHMFL, which is supported by NSF DMR-1157490 and the State of Florida. Work at NRL was supported by core programs and the NRL Nanoscience Institute, and by AFOSR under contract number AOARD 14IOA018-134141. Work at Rice University was supported by the AFOSR under FA9550-14-1-0268.


\begin{references}

\bibitem{Miura}N. Miura, \emph{Physics of Semiconductors in High Magnetic Fields}, Oxford University Press, (2008).

\bibitem{Knox}R. S. Knox, \emph{Theory of Excitons}, Academic Press, New York, (1963).

\bibitem{Splendiani}A. Splendiani \emph{et al.}, Nano Lett. \textbf{10}, 1271 (2010).

\bibitem{Mak2010}K. F. Mak, C. Lee, J. Hone, J. Shan, and T. F. Heinz, Phys. Rev. Lett. \textbf{105}, 136805 (2010).

\bibitem{Xiao}D. Xiao, G.-B. Liu, W. Feng, X. Xu, and W. Yao, Phys. Rev. Lett. \textbf{108}, 196802 (2012).

\bibitem{XuReview}X. Xu, W. Yao, D. Xiao, and T. F. Heinz, Nature Phys. \textbf{10}, 343 (2014).

\bibitem{MakNatNano}K. F. Mak, K. He, J. Shan, and T. F. Heinz, Nature Nanotech. \textbf{7}, 494 (2012).

\bibitem{Zeng}H. Zeng, J. Dai, W. Yao, D. Xiao, and X. Cui, Nature Nanotech. \textbf{7}, 490 (2012).

\bibitem{Sallen}G. Sallen \emph{et al.}, Phys. Rev. B \textbf{86}, 081301(R) (2012).

\bibitem{Cao}T. Cao \emph{et al.}, Nat. Commun. \textbf{3}, 887 (2012).

\bibitem{MacNeill}D. MacNeill \emph{et al.}, Phys. Rev. Lett. \textbf{114}, 037401 (2015).

\bibitem{Srivastava}A. Srivastava, M. Sidler, A. V. Allain, D. S. Lembke, A. Kis, and A. Imamo\u{g}lu, Nature Phys. \textbf{11}, 141 (2015).

\bibitem{Aivazian}G. Aivazian \emph{et al.}, Nature Phys. \textbf{11}, 148 (2015).

\bibitem{Li}Y. Li \emph{et al.}, Phys. Rev. Lett. \textbf{113}, 266804 (2014).

\bibitem{Wang}G. Wang \emph{et al.}, 2D Materials \textbf{2}, 034002 (2015).

\bibitem{Mitioglu}A. A. Mitioglu \emph{et al.}, Nano Lett. \textbf{15}, 4387 (2015).

\bibitem{Yang1}L. Yang \emph{et al.}, Nature Nanotech. \textbf{11}, 830 (2015).

\bibitem{Yang2}L. Yang \emph{et al.}, Nano Lett. \textbf{15}, 8250 (2015).

\bibitem{Stier}A. V. Stier, K. M. McCreary, B. T. Jonker, J. Kono, and S. A. Crooker, Nat. Comm. \textbf{7}, 10643 (2016).

\bibitem{Ram}A. Ramasubramaniam, Phys. Rev. B \textbf{86}, 115409 (2012).

\bibitem{Komsa} H.-P. Komsa, A. V. Krasheninnikov, Phys. Rev. B \textbf{86}, 241201 (2012). 

\bibitem{Liu}G.-B. Liu, W.-Y. Shan, Y. Yao, W. Yao, D. Xiao, Phys. Rev. B \textbf{88}, 085433 (2013). 

\bibitem{Ye}Z. Ye \emph{et al.}, Nature \textbf{513}, 214 (2014). 

\bibitem{KormanyosPRX}A. Korm\'{a}nyos, V. Z\'{o}lyomi, N. D. Drummond, and G. Burkard, Phys. Rev. X \textbf{4}, 011034 (2014). 

\bibitem{Chernikov} A. Chernikov \emph{et al.}, Phys. Rev. Lett. \textbf{113}, 076802 (2014).

\bibitem{He}K. He \emph{et al.}, Phys. Rev. Lett. (\textbf{113}), 026803 (2014). 

\bibitem{Zhu} B. Zhu, X. Chen, X. Cui, Scientific Reports \textbf{5}, 9218 (2015).

\bibitem{WangPRL2015}G. Wang \emph{et al.}, Phys. Rev. Lett. \textbf{114}, 097403 (2015). 

\bibitem{Stroucken}T. Stroucken and S. W. Koch, J. Phys. Condens. Matter \textbf{27}, 345003 (2015).

\bibitem{Hanbicki}A. Hanbicki \emph{et al.}, Solid State Commun. \textbf{203}, 16 (2015).

\bibitem{McCreary}K. M. McCreary, A. T. Hanbicki,  G. G. Jernigan, J. C. Culbertson, and B. T. Jonker, Sci. Rep. \textbf{6}, 19159 (2016).

\bibitem{Crooker1}S. A. Crooker, D. G. Rickel, S. K. Lyo, N. Samarth, D. D. Awschalom, Phys. Rev. B \textbf{60}, 2173 (1999).

\bibitem{Crooker2}S. A. Crooker, E. Johnston-Halperin, D. D. Awschalom, R. Knobel, N. Samarth, Phys. Rev. B \textbf{61}, 16307 (2000).
\bibitem{Astakhov}G. V. Astakhov \emph{et al.}, Phys. Rev. B \textbf{71}, 201312 (2005).

\bibitem{Alberi}K. Alberi \emph{et al.}, Phys. Rev. Lett. \textbf{110}, 156405 (2013).

\bibitem{Zeke}E. Johnston-Halperin \emph{et al.}, Phys. Rev. B \textbf{63}, 205309 (2001).

\bibitem{Furis}M. Furis, J. A. Hollingsworth, V. I. Klimov, and S. A. Crooker, J. Phys. Chem. B \textbf{109}, 15332 (2005).

\bibitem{Zaric}S. Zaric \emph{et al.}, Phys. Rev. Lett. \textbf{96}, 016406 (2006).

\bibitem{Shaver}J. Shaver \emph{et al.}, Phys. Rev. B \textbf{78}, 081402 (2008).

\bibitem{Diaconu}C. V. Diaconu \emph{et al.}, J. Appl. Phys. \textbf{109}, 073513 (2011).

\bibitem{Walck}S. N. Walck and T. L. Reinecke, Phys. Rev. B \textbf{57}, 9088 (1988).

\bibitem{Keldysh}L. V. Keldysh, JETP Lett. \textbf{29}, 658 (1979).

\bibitem{Cudazzo}P. Cudazzo, I. V. Tokatly, and A. Rubio,  Phys. Rev. B \textbf{84}, 085406 (2011).

\bibitem{Berkelbach}T. C. Berkelbach, M. S. Hybertsen, and D. R. Reichman, Phys. Rev. B \textbf{88}, 045318 (2013).












\end{references}
\end{document}